\documentclass[aps,pre,groupedaddress,10pt]{revtex4-1}

\usepackage[active]{srcltx}
\usepackage{graphicx}
\usepackage{amsmath, amssymb, amsfonts}
\usepackage{color}

\begin{document}


\title{Susceptible-Infected-Susceptible dynamics with mitigation in connection of infected population}


\author{K. M. Kim$^{1}$, C. Dias$^{1}$ and M. O. Hase$^{1}$}
\thanks{The authors wish to congratulate Prof. S. R. Salinas on the occasion of his $80^{\text{th}}$ birthday in 2022.}
\affiliation{$^{1}$ Escola de Artes, Ci\^encias e Humanidades, Universidade de S\~ao Paulo, Av. Arlindo B\'ettio 1000, 03828-000 S\~ao Paulo, Brazil}


  
\begin{abstract}
\begin{center}
\textbf{\large Abstract}
\end{center}
  
The susceptible-infected-susceptible epidemic model is analyzed through a degree-based mean-field approach. In this work, a mitigation factor is introduced in the probability of finding an infected individual following an edge. This modification simulates situations where the infected population reduces its participation in the dynamics of disease propagation, as may happen with the seclusion or hospitalization of infected individuals. A detailed investigation of this new model and its comparison to the original one (without the mitigation factor) was performed on the Barab\'asi-Albert network, where some important results were analytically accessible.

\medskip
Keywords: Nonequilibrium statistical physics; networks; mathematical modeling in epidemiology
\end{abstract}

\pacs{}
\email{Corresponding author: mhase@usp.br}


\keywords{}

\maketitle



\section{Introduction}
\label{sec:introduction}

Infectious diseases have been a common cause of mortality over the years, especially in low-income countries \cite{WHO}. However, after the recent outbreak of COVID-19, it became clear that infectious diseases can emerge and disseminate on a more global scale. At the same time, forecasting the way a disease spreads over the population is a challenging problem. The difficulty to make predictions on the course of this propagation based on incomplete data (as a consequence of underreporting, for example) and the inviability of performing experiments on infectious disease proliferation in the human population justify the mathematical modeling as a powerful approach to face these problems \cite{H89}.

The mathematical modeling in epidemiology usually divides the population into compartments, which characterizes the states of individuals with respect to disease: they can be susceptible, infected, exposed, and so on. The main idea is to write a set of differential equations that specifies the flow of individuals between these compartments \cite{TdO20}. One important issue here is to be able to understand the network of interaction between individuals when the infection is propagated by direct contact between them. The traditional approach relies on the \textquotedblleft homogeneous mixing\textquotedblright \cite{H06}, in which any individual has the same probability of interacting with any other member of a given compartment. Although this strategy has been widely adopted over decades \cite{B75}, it clearly diverges from the heterogeneous organization expected in human contacts \cite{DM03}. This means that a theory that combines both classical mathematical epidemiology \cite{AM92, M02} and network theory \cite{AB02, N10, dM20, DM22} is required to enhance our understanding of disease propagation in a more realistic setup \cite{FCPS12, PSCvMV15}.

In this work, we concentrate on a common aspect observable in epidemics. As expected, part of infected individuals tend to weak their ties to other ones by seclusion or deliberate isolations (like hospitalizations), and this means that the network of contact between members of the system is changed. In a recent work \cite{DH21}, this effect was incorporated in the dynamics through a tunable parameter that controls the probability of linking between individuals. Here, we take a different approach but based on the hypothesis that the probability of hitting an infected individual should, at some point, decrease as the infected population grows. The simplest way of introducing this effect goes back to the classical work of Verhulst \cite{V38}, as explained below.

In population dynamics, the Malthus model \cite{M98} states that the variation of a population is proportional to its size. In this scenario, the population grows exponentially without any constraint and under the hypothesis of having an infinite amount of resources that support this prosperity. To avoid this uncontrolled growth, the introduction of a carrying capacity that fixes the maximum population of the system was proposed by Verhulst; specifically, the variation of a population was replaced by a parabola with negative convexity instead of a line (Malthus model). This simple idea has far-reaching consequences, and the population growth in this new setup follows now a logistic function. The same idea was already tested in the context of networks \cite{HCG16}, and we are now applying it to describe the interplay between individuals where an epidemic takes place.

The layout of this work is as follows. In section 2, we review the heterogeneous mean-field (HMF) model analyzing its criticality and prevalence. In section 3, a modified version of the HMF is presented and investigated, and the conclusions are summarized in the last section.


\section{Susceptible-Infected-Susceptible dynamics on heterogeneous mean-field model}
\label{sec:sis}

We will investigate the susceptible-infected-susceptible (SIS) dynamics \cite{R10} in this work. According to this model, the population is divided into infected and susceptible ones and two discrete events take place: (i) an infected individual can infect a susceptible member of the system by contact or (ii) an infected individual recovers spontaneously and joins the set of susceptible ones. The mean-field approximation of this dynamics predicts two different scenarios separated by a positive critical point \cite{MD99}. If the infection rate is sufficiently high, the model converges to an endemic state, where susceptible and infected individuals coexist. On the other hand, in the subcritical case, the disease dies out.

The first mean-field scheme proposed for the SIS dynamics that can deal with heterogeneities of the underlying network is presented in this section. This is the so-called heterogeneous mean-field model (HMF) \cite{PSV01a, PSV01b}  -- or degree-based mean-field --, and captures the structure of the network through its degree distribution. In this model, the degree is considered an annealed variable instead of a quenched one -- the latter approach is also a popular one and leads, in general, to different results (see, for instance, \cite{HM08, CH16}), but will not be treated here. The original formulation of this model is reviewed here for two reasons: first, we will introduce some notations and concepts; second, the main strategy to tackle this system is valuable to analyze our proposed model, which will be shown later.


\subsection{Formulation of the heterogeneous mean-field model}
\label{subsec:hmf_def}

In this formulation, vertices are partitioned according to their degrees: nodes that share the same degree are considered to have the same statistical properties. The key quantity, therefore, is $\rho_{k}(t)$, which is the probability of a vertex with degree $k$ being infected (or the fraction of infected individuals that have degree $k$) at time $t$. The master equation describing its time evolution \cite{PSV01a} is given by
\begin{align}
\frac{\partial}{\partial t}\rho_{k}(t) = -\rho_{k}(t) + \lambda k\big[1-\rho_{k}(t)\big]\Theta_{k}(t),
\label{hmf}
\end{align}
where the recovering rate was taken to be $1$ without loss of generality, as it is always possible by a suitable time scaling. Here, $\lambda$ is the infection rate and $\Theta_{k}(t)$ is the probability that a link emerging from a vertex of degree $k$ connects to an infected node.

In this work, we are mostly interested in the stationary state, in which $\rho_{k}(t)=\rho_{k}$ becomes a time-independent function. In this case, it is straightforward that $\frac{\partial\rho_{k}}{\partial t}=0$, and this leads to
\begin{align}
\rho_{k} = \frac{\lambda k\Theta_{k}}{1+\lambda k\Theta_{k}},
\label{hmf_st}
\end{align}
where $\Theta_{k}(t)=\Theta_{k}$ is also a stationary probability.

Denoting by $P(k^{\prime}|k)$ the conditional probability that a vertex of degree $k$ links to another with degree $k^{\prime}$, one seees that
\begin{align}
\Theta_{k} = \sum_{k^{\prime}}P(k^{\prime}|k)\rho_{k^{\prime}}.
\label{Theta_k}
\end{align}

In this work, however, we will consider uncorrelated networks, \textit{id est}, the probability that a vertex of degree $k$ links to another with degree $k^{\prime}$ does not depend on $k$. Then, the conditional probability $P(k|\ell)$ is reduced to $P(k|\ell)=q(k)$, which stands for the probability of connecting to a node of degree $k$, and is given by \cite{DM03}
\begin{align}
q(k) = \frac{kP(k)}{\langle k\rangle},
\label{q}
\end{align}
where $P$ is the degree distribution of the network and $\langle k^{n}\rangle := \sum_{k}k^{n}P(k)$ stands for the $n$-th moment. Then, in this setup the probability $\Theta_{k}$ is actually independent on $k$, and it will henceforth be denoted by $\Theta_{k}\equiv\Theta$, where
\begin{align}
\Theta = \sum_{k}q(k)\rho_{k} = \frac{1}{\langle k\rangle}\sum_{k}kP(k)\rho_{k}
\label{Theta}
\end{align}
after \eqref{q}.

The main strategy to access the important properties of this stationary system is the analysis of \eqref{hmf_st} and \eqref{Theta}. It can also characterize the prevalence
\begin{align}
\rho = \sum_{k}\rho_{k}P(k),
\label{rho}
\end{align}
which is the stationary infection probability.


\subsection{HMF model on Barab\'asi-Albert network}
\label{subsec:hmf}

The equations above constitute the main setup for the heterogeneous mean-field approach for the SIS model. In this work, the dynamics will be examined on a Barab\'asi-Albert (BA) model. Under the continuous approximation (the degree is assumed, for convenience, to be a continuous variable), the degree distribution is given by
\begin{align}
P(k) = \frac{2m^{2}}{k^{3}} \qquad (k\geq m),
\label{P_BA}
\end{align}
where $m$ is the minimum degree of the system. One can also check that the mean degree is $\langle k\rangle=2m$. From \eqref{hmf_st}, \eqref{Theta} and \eqref{P_BA}, and taking the continuous limit, it is straightforward that
\begin{align}
\Theta = \lambda m\Theta\ln\left(1+\frac{1}{\lambda m\Theta}\right).
\label{eqTheta_hmf}
\end{align}
This relation displays a trivial solution, $\Theta=0$, which stands for the absorbent state where no infected individual is present. If we search for a nontrivial solution ($\Theta\neq 0$), then \eqref{eqTheta_hmf} leads to
\begin{align}
\Theta = \frac{1}{\lambda m}\left(\frac{1}{e^{\frac{1}{\lambda m}}-1}\right),
\label{nontrivialTheta_hmf}
\end{align}
which is an increasing function with respect to $\lambda m$, as one can see from
\begin{align}
\frac{\textup{d}}{\textup{d}(\lambda m)}\Theta = \frac{e^{\frac{1}{\lambda m}} - \lambda m\big(e^{\frac{1}{\lambda m}}-1\big)}{\big(\lambda m\big)^{3}\big(e^{\frac{1}{\lambda m}}-1\big)^{2}},
\label{derivativeTheta_hmf}
\end{align}
which can be shown to be positive.
In addition to this property, one can also see that the probability $\Theta$ is a function of the product $\lambda m$ and
\begin{align}
\lim_{\lambda\rightarrow\infty}\Theta = 1.
\label{limTheta1_hmf}
\end{align}

The probability \eqref{nontrivialTheta_hmf} allows one to obtain the stationary infection probability of vertices with degree $k$ through \eqref{hmf_st}. On BA model, the prevalence \eqref{rho} is then
\begin{align}
\rho = 2\left(\lambda m\Theta\right)^{2}\left[ \frac{1}{\lambda m\Theta} - \ln\left(1 + \frac{1}{\lambda m\Theta}\right)  \right],
\label{rho_hmfBA}
\end{align}
where $\Theta$ is given by \eqref{nontrivialTheta_hmf}. As one can see, the prevalence is positive for any positive infection rate $\lambda$ and vanishes when $\lambda=0$ only. This means that the infection is persistent in this model and the infection threshold is $\lambda_{c}=0$. In the neighborhood of this critical point, the prevalence behaves asymptotically as
\begin{align}
\rho \simeq 2e^{-\frac{1}{\lambda m}}.
\label{rho_hmfBA_lambda0}
\end{align}
In section \ref{subsec:prevalence}, we will also show that the prevalence \eqref{rho} is a monotonically increasing function with respect to the infection rate $\lambda$.


\section{SIS dynamics on modified HMF model}
\label{sec:mHMF}

In this section, a modified version of the HMF model (mHMF) is introduced. The starting point is the equation \eqref{Theta}, which relates the probabilities $\Theta$ and $\rho_{k}$ on an uncorrelated network. Here, this relation is replaced by
\begin{align}
\Theta = \sum_{k}q(k)\rho_{k}\left(1-\rho_{k}\right) = \frac{1}{\langle k\rangle}\sum_{k}kP(k)\rho_{k}\left(1-\rho_{k}\right).
\label{Theta_mhmf}
\end{align}
Comparing this proposal to \eqref{Theta}, a mitigating factor is attached to each term in the summation. This is a simple change in the same spirit of the classical Malthus-Verhulst model \cite{V38}, where the population growth is weakened by a factor similar to the one inserted in \eqref{Theta_mhmf}. In other words, the factor $(1-\rho_{k})$ prevents the increase of the probability $\Theta$ as $\rho_{k}$ increases. This choice has some obvious motivations, as infected individuals may isolate themselves (by resting or taking protective measures) and retreat from the infection dynamics. Therefore, we are introducing a modification that decreases the probability of meeting an infected individual when the infected population becomes large. At the same time, this change is less perceptible when the population of infected individuals is not significant -- a scenario where the system does not \textquotedblleft feel\textquotedblright the impact of the infection due to the few cases.

The first important consequence is that the maximum probability $\Theta$ is now less than $1$. Although its value is unimportant here, we will see the impact of introducing such mitigating factors on the behavior of this probability and analyze the implication on the prevalence.


\subsection{Infection threshold of mHMF}
\label{subsec:criticality}

The main relation \eqref{Theta_mhmf} can be cast as
\begin{align}
\Theta = \lambda m\Theta\left[ \ln\left(1+\frac{1}{\lambda m\Theta}\right) - \frac{1}{1+\lambda m\Theta} \right]
\label{eqTheta_mhmf}
\end{align}
on a BA network. As before, the trivial solution $\Theta=0$ corresponding to the absorbent phase is also present. Let us now search for a nontrivial solution $(\Theta\neq 0)$ for this relation, that can then be rewritten as
\begin{align}
g(\Theta) = 1,
\label{a2}
\end{align}
where
\begin{align}
g(\Theta) = \lambda m\left[ \ln\left(1+\frac{1}{\lambda m\Theta}\right) - \frac{1}{1+\lambda m\Theta} \right].
\label{a3}
\end{align}
Searching for a solution of (\ref{a2}) is equivalent to the problem of finding an intersection point of the curves $y=g(\Theta)$ and the line $y=1$ in the coordinate system $y\times\Theta$. Since the derivative of $g$ (with respect to $\Theta$),
\begin{align}
g^{\prime}(\Theta) = -\frac{\lambda m}{\Theta\left(\lambda m\Theta+1\right)^{2}},
\label{a5}
\end{align}
is negative, $g$ is a decreasing function. Furthermore, in the extremal points, we have $\lim_{\Theta\rightarrow 0^{+}}g(\Theta)=+\infty$ and
\begin{align}
g(\Theta=1) = \lambda m\ln\left(1+\frac{1}{\lambda m}\right) - \frac{\lambda m}{1+\lambda m}.
\label{a6}
\end{align}

\begin{figure}
\begin{center}
\includegraphics[width=246pt]{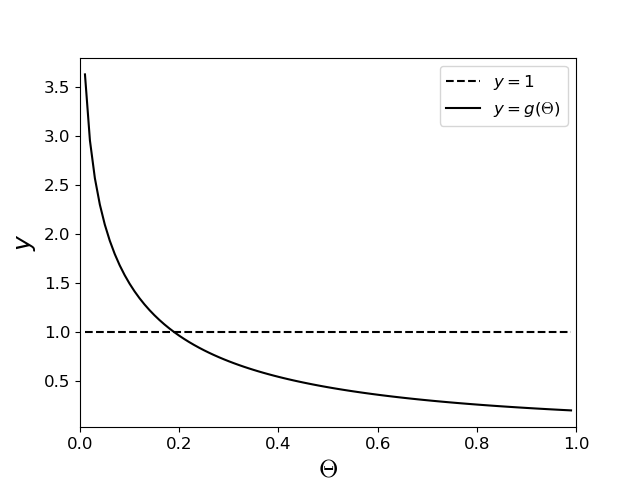}
\caption{\label{yg} Graphs $y=1$ and $y=g(\Theta)$; here, $\lambda m=1$.}
\end{center}
\end{figure}

Then, it remains to verify if $g(\Theta=1)$, as a function of $\lambda$, intercepts the line $y=1$ as in figure \ref{yg}. If $g(1)<1$, the equation \eqref{a2} has always a solution, which means that the infection is always present; in other words, we would have $\lambda_{c}=0$. We will show that it is, indeed, the case. To see this result, note that
\begin{align}
g(1) = \lambda m\ln\left(1+\frac{1}{\lambda m}\right) - \frac{\lambda m}{1+\lambda m} < \lambda m\ln\left(1+\frac{1}{\lambda m}\right),
\label{a7}
\end{align}
since $\lambda,m>0$. Since the function $\Psi$, defined by
\begin{align}
\Psi(z) = z\ln\left(1+\frac{1}{z}\right),
\label{a8}
\end{align}
is increasing (note that the last term of \eqref{a7} is $\Psi(\lambda m)$) and $\lim_{z\rightarrow\infty}\Psi(z) = 1$, one has $g(1)<\Psi(\lambda m)\leq\Psi(\infty)=1$, which completes the proof.

Therefore, the nontrivial solution of \eqref{eqTheta_mhmf} is always present. In the next section, we will examine the prevalence and its behavior when the infectious rate is small.


\subsection{Prevalence of mHMF}
\label{subsec:prevalence}

\begin{figure}
\begin{center}
\includegraphics[width=246pt]{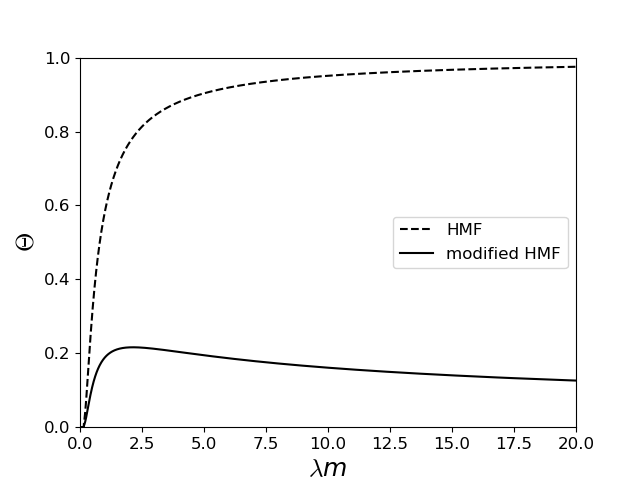}
\caption{\label{thetafig} Graph $\Theta\times\lambda m$ for HMF and modified HMF models.}
\end{center}
\end{figure}

As in the HMF, the probability $\Theta$ is also a function of the product $\lambda m$, but it displays a different behavior in mHMF, as we can see in figure \ref{thetafig}, which is a numerical solution of \eqref{eqTheta_hmf} and \eqref{eqTheta_mhmf}. This function is not monotonically increasing as in the HMF case and a peak can be detected in this model at the point $\lambda_{p}m$. From \eqref{a2} and imposing $\frac{\partial g}{\partial\lambda}=0$, one can see that
\begin{align}
1+\sqrt{\lambda_{p}m} = \lambda_{p}m\ln\left(\frac{\sqrt{\lambda_{p}m}}{\sqrt{\lambda_{p}m}-1}\right).
\label{peak}
\end{align}
At this maximum point, the corresponding probability $\Theta_{p}$ is
\begin{align}
\Theta_{p} = \frac{\sqrt{\lambda_{p}m}-1}{\lambda_{p}m}.
\label{Thetapeak}
\end{align}

\begin{figure}
\begin{center}
\includegraphics[width=246pt]{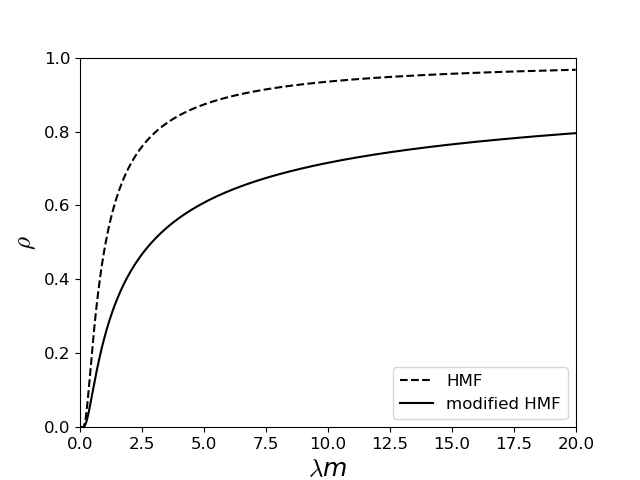}
\caption{\label{rhofig} Graph $\rho\times\lambda m$ for HMF and modified HMF models.}
\end{center}
\end{figure}

The prevalence of the mHMF is given by
\begin{align}
\rho = 2\left(\lambda m\Theta\right)^{2}\left[\frac{1}{\lambda m\Theta}-\ln\left(1+\frac{1}{\lambda m\Theta}\right)\right].
\label{rho_mhmf}
\end{align}
Although the dependence of this quantity to $\Theta$ is the same as HMF, both prevalences do not match since $\Theta$ are different. In HMF, we have seen that the probability $\Theta$ of finding an infected vertex following an edge is monotonically increasing with the infectious rate $\lambda$, but in mHMF, it displays a maximum value. We will now see that despite this difference in both models, the prevalence is monotonically increasing in both cases, as shown in figure \ref{rhofig}, which is the numerical solution of both models. Let us analyze the function
\begin{align}
\frac{\textup{d}}{\textup{d}\lambda}\rho = 2m\left(\Theta+\lambda\frac{\textup{d}}{\textup{d}\lambda}\Theta\right)\Psi_{2}(\lambda m\Theta)
\label{drho_dlambda}
\end{align}
with
\begin{align}
\Psi_{2}(x) := 2 - 2x\ln\left(1 + \frac{1}{x}\right) - \frac{1}{1 + x} \qquad (x>0).
\label{Psi2}
\end{align}
We will now see that the function \eqref{drho_dlambda} is positive for both HMF and mHMF models. The strategy consists in showing that both the factor $\Theta+\lambda\frac{\textup{d}}{\textup{d}\lambda}\Theta$ and $\Psi_{2}$ are positive (this is a sufficient condition, although not necessary). Evaluating the derivative of $\Psi_{2}$,
\begin{align}
\Psi_{2}^{\prime}(x) = -2\ln\left(1+\frac{1}{x}\right) + \frac{2}{x+1} + \frac{1}{\left(x+1\right)^{2}},
\label{dPsi2}
\end{align}
and its second derivative,
\begin{align}
\left(\Psi_{2}^{\prime}\right)^{\prime}(x) = \frac{2}{x\left(x+1\right)^{3}} > 0,
\label{ddPsi2}
\end{align}
one sees that the function $\Psi_{2}^{\prime}(x)$ is a monotonically increasing function, which means that it reaches its supremum at $x\rightarrow\infty$. Considering that $\lim_{x\rightarrow\infty}\Psi_{2}^{\prime}(x)=0$, $\Psi_{2}^{\prime}$ is then negative, which implies that $\Psi_{2}$ is a decreasing function. Therefore, $\Psi_{2}$ is minimum at $x\rightarrow\infty$, and as $\lim_{x\rightarrow\infty}\Psi_{2}(x)=0$, one sees that $\Psi_{2}$ is non-negative.

In the HMF model, we have seen that \eqref{derivativeTheta_hmf} is a positive function, which leads to $\Theta+\lambda\frac{\textup{d}}{\textup{d}\lambda}\Theta>0$, which justifies the positivity of the derivative of the prevalence for that model. On the other hand, the probability $\Theta$ of finding an infected node following an edge is not monotonically increasing in the mHMF model. From \eqref{a2} and \eqref{a3}, one can evaluate the derivative $\frac{\textup{d}}{\textup{d}\lambda}\Theta$, and show that
\begin{align}
\Theta + \lambda\frac{\textup{d}}{\textup{d}\lambda}\Theta = \frac{1}{\left(\lambda m\right)^{2}}\frac{1}{\left[\frac{1}{\lambda m\Theta}-\frac{1}{\left(1+\lambda m\Theta\right)^{2}}\right]} > 0.
\label{dTheta_mhmf}
\end{align}
Hence, the prevalence is also monotonically increasing for the mHMF model.

\section{Conclusion}
\label{conclusion}

In this work, a modified version of the heterogeneous mean-field that introduces mitigation in the participation of infected nodes in the infection dynamics was proposed. This proposal is motivated by the possibility of infected individuals retreating (at least partially) from the propagation dynamics and goes back to the classical work of Verhulst. This idea was tested on the Barab\'asi-Albert network, which is scale-free and, at the same time, many results are analytically accessible.

The main difference is the drastic change in the profile of the probability of an edge finding an infected node, which displays a peak. In the original HMF, it is a monotonically increasing function. Nevertheless, the prevalence of both models is increasing with respect to the infection rate -- although quantitative differences do exist. We managed to support all these conclusions based on exact results.


\section{Acknowledgements}

The authors thank A. S. da Mata for fruitful observations.


\section{Declarations}

Conflict of interests: The authors declare no competing interests.

\end{document}